# Superconductivity in the C32 Intermetallic Compounds $AAl_{2-x}Si_x$, with A = Ca and Sr; and $0.6 < x < 1.2$

B. Lorenz[1], J. Lenzi[1], J. Cmaidalka[1], R. L. Meng[1], Y. Y. Sun[1], Y. Y. Xue[1], C. W. Chu[1,2,3]

[1] Texas Center for Superconductivity and Department of Physics, University of Houston, Houston, TX 77204-5932

[2] Lawrence Berkeley National Laboratory, 1 Cyclotron Road, Berkeley, CA 94720

[3] Hong Kong University of Science and Technology, Hong Kong, China

**Abstract**

The intermetallic compounds $AAl_{2-x}Si_x$, where A = Ca, Sr or Ba, crystallize in the C32 structure, same as the recently discovered $MgB_2$ with a high superconducting transition temperature of 39 K. For x = 1, superconductivity has been observed in AAlSi with A = Ca and Sr, but not with A = Ba. The transition temperatures are 7.8 and 5.1 K, respectively for CaAlSi and SrAlSi. The $CaAl_{2-x}Si_x$ compound system display a $T_c$-peak at x = 1, a possible x-induced electronic transition at x ~ 0.75 and a possible miscibility gap near x ~ 1.1 which results in a very broad superconducting transition. The Seebeck coefficients of AAlSi indicate that their carriers are predominantly electrons in nature, in contrast to the holes in $MgB_2$.

## 1. Introduction

The discovery of superconductivity at 39 K in $MgB_2$ [1] has rekindled the interest in intermetallic compounds. The specific structure of $MgB_2$ with boron forming a perfect honeycomb lattice and magnesium filling the spaces between the boron planes has drawn the interest into other binary and pseudo-ternary compounds crystallizing in the same C32 structure as $MgB_2$. Superconductivity has recently been found in $Sr(Ga_{0.37}Si_{0.63})_2$ at 3.4 K [2] and CaAlSi at 7.7 K [3]. These compounds are isostructural to $MgB_2$ since the ions in the honeycomb planes (Ga/Si or Al/Si) are randomly distributed and do not form a superstructure. According to early reports [4, 5] the $CaAl_{2-x}Si_x$ crystallizes in the C32 structure for $0.65 \leq x \leq 1.5$. This wide range of structural stability and the recently reported observation of superconductivity for x=1 in CaAlSi opens the interesting possibility to change the electronic structure by varying x (i.e. the Al/Si ratio) and investigate its effect on the superconducting state. Besides the change of the average valence with x, a lattice compression is expected if the larger ion Al is substituted by the smaller Si in the honeycomb planes. If the larger Sr and Ba replace the isoelectronic smaller Ca a lattice expansion is expected along both the c- and a- directions. In order to study the effects of valence change and lattice expansion on the superconducting properties of the C32 intermetallic compounds, we have synthesized AAlSi for A = Ca, Sr and Ba as well as the pseudo-ternary $CaAl_{2-x}Si_x$ polycrystalline samples with $0.6 \leq x \leq 1.2$. The samples are characterized by x-ray, resistivity, and Seebeck coefficient measurements. Sharp x-ray diffraction patterns were evident for the stoichiometric AAlSi, i.e. for the x = 1 compounds, suggesting the possible Al/Si-order in the compounds. Superconductivity was detected in CaAlSi at 7.8 K, consistent with previous report; and in SrAlSi at 5.1 K. The transition temperature $T_c$ of $CaAl_{2-x}Si_x$ peaks at x = 1 and deceases rapidly as x deviates from 1. In contrast to the isostructural $MgB_2$, the Seebeck coefficient of CaAlSi and SrAlSi indicates predominantly n-type conduction. Such difference in the electronic structure may be responsible for the large difference in $T_c$ between $MgB_2$ and other isostructural intermetallic compounds.

## 2. Sample Preparation and Structural Analysis

The pseudo-ternary intermetallic compounds AAlSi for A = Ca, Sr, and Ba; and $CaAl_{2-x}Si_x$ ($0.6 \leq x \leq 1.2$) have been synthesized by argon arc melting with appropriate amounts of Ca (99 %), Sr (99 %), Ba (99.5%), Al (99.99 %), and Si (99.99 %). The weight loss during the arc melting process was less than 1 %. The structure of the samples was examined by powder x-ray diffraction. All AAlSi show one main phase with the C32 hexagonal structure and only minor trace of unidentified impurities. However, the x-ray patterns of $CaAl_{2-x}Si_x$ also show the presence of a trace of $Al_2Si$ for x between 1 and 1.2. When x decreases below 0.7, there are three minor impurity phases, $Al_2Si$, $Al_4Si$, and $CaAl_2Si_2$ indicating the limit of Si –doping of the compound. The relatively high phase purity of AAlSi in comparison with the $AAl_{2-x}Si_x$ compounds with $x \neq 1$ suggests that atomic ordering may have taken place in these x = 1 compounds.

The lattice parameters of the hexagonal unit cell of $CaAl_{2-x}Si_x$ decrease with the Si content (x). Fig. 1 displays both lattice constants, a and c, as a function of x. The relative compression of the lattice in the hexagonal plane is about 60 % larger than that perpendicular to the planes, i. e. dlna/dx = -0.028, and dlnc/dx = -0.018. This is not surprising because the smaller Si ions occupy the sites within the honeycomb planes.

Both the lattice compression and the change of the average valence should affect the superconducting properties in this compound.

In contrast, the lattices of SrAlSi and BaAlSi are expanded more in the c-direction than in the a-direction as compared with CaAlSi. The values of a and c for AAlSi are (a, c) = (4.189 Å, 4.400 Å), (4.220 Å, 4.754 Å) and (4.290 Å, 5.140 Å) for A = Ca, Sr, and Ba, respectively. The expansion is expected since the larger Sr or Ba- ions replace the smaller Ca-ions in between the honeycomb planes.

## 3. Transport Properties and Superconductivity

Because of the structural similarity of AAlSi and $MgB_2$ it appears to be interesting to compare the electronic structure and transport properties. Four lead resistivity measurements have been conducted employing the low frequency (19 Hz) ac resistance bridge LR 700 (Linear Research). For measuring the Seebeck coefficient a home made highly sensitive ac technique was used which allows to estimate the Seebeck coefficient within a precision of less than 0.1 μV/K. The details of this setup have been described elsewhere [6].

The Seebeck coefficient, S(T), may provide substantial information about the nature of carriers at the Fermi surface. For $MgB_2$ the hole character of the charge carriers was first concluded from the positive Seebeck coefficient [7,8] and later confirmed by Hall effect measurements [9]. Since BaAlSi is not superconducting, we have examined only CaAlSi and SrAlSi. The Seebeck coefficient for CaAlSi is shown in Fig. 2a. S(T) is negative over the whole temperature range and drops to zero at $T_c \approx 7.8$ K (see upper right inset). The drop of the resistivity at $T_c$ is shown in the lower left inset of Fig. 2a and is similar as reported before for CaAlSi [3]. A small, shoulder like feature just above $T_c$ probably indicates a contribution from the phonon drag to S(T). Similar data have been obtained for SrAlSi (Fig. 2b). S(T) is negative for all temperatures and exhibits a small, but distinct drop to zero at about 5.1 K (upper right inset in Fig. 2b). The resistivity data (lower left inset) clearly show the onset of superconductivity at this critical temperature. This is, to our knowledge, the first observation of superconductivity in the intermetallic compound SrAlSi. The phonon drag contribution to S(T) appears to be more pronounced in SrAlSi than in CaAlSi.

As compared to $MgB_2$, the Seebeck coefficient of AAlSi is very different, revealing major differences in the electronic structure of both compounds. So it is not surprising that the superconducting transition temperature is actually far lower than that of $MgB_2$. From the negative sign of S(T) it is concluded that the predominant carriers are electrons instead of holes like in $MgB_2$. It is interesting to point out that the electron-doped cuprates generally exhibit a lower $T_c$ than the hole-doped ones. However, this conclusion needs to be confirmed by Hall effect measurements. In addition the Al and Si ions occupying the honeycomb planes are far heavier than the boron ions giving rise to a lower value of the in-plane phonon frequencies. Even if the mechanism for superconductivity in CaAlSi or SrAlSi would be very similar to the one discussed in $MgB_2$, a lower $T_c$ is to be expected. It should be noted that the relatively high $T_c$ of $MgB_2$ was attributed in part to the high in-plane phonon frequency.

The pseudo-ternary compound series $CaAl_{2-x}Si_x$, however, is of special interest since it forms the stable C32 structure within certain limits of x. This opens the possibility to tune the electronic structure by slightly changing the composition in $CaAl_{2-x}Si_x$ and to observe the effect on superconductivity. The resistivity data near the superconducting transition for $0.6 \leq x \leq 1$ are shown in Fig. 3. The sharp resistive superconducting transition shifts almost parallel to lower temperature if x decreases from 1 to 0.8. Below x=0.8 the transition starts broadening and no clear tendency of $T_c$ as a function of x is further observed. It should be noted that the x-ray spectra for these compositions (x<0.8) indicate an increasing amount of second phases such as $Al_2Ca$ and $Al_4Ca$. The $T_c$ values obtained are summarized in Fig. 4. The error bars attached to the data in Fig. 4 indicate the width of the transitions as estimated from a 90 % to 10 % resistivity change at $T_c$. The $T_c$ clearly drops off rapidly as x deviates from 1 with an accompanying increase in the transition width, suggesting a possible ordering at x = 1, consistent with the x-ray data. $T_c$ remains almost constant as x is reduced to below 0.75. The drastic change in $T_c$ at x ~ 0.75 may reflect a change of the density of states at the Fermi energy with decreasing x similar to the sharp $T_c$-dip associated with a possible doping-induced electronic transition observed at x = 1.3 in the isoelectronic and isostructural $CaGa_{2-x}Si_x$ series [10]. However, detailed study is needed to determine the existence and nature of the proposed electronic transition, especially in view of the presence of increasing amounts of the impurity phases with x-decrease to below 0.75. The transition width for $CaAl_{0.9}Si_{1.1}$ is unusually large. It indicates a possible non-uniformity of the Al distribution throughout the sample. The resistivity transition actually proceeds in two broad steps (Fig. 5). We have tried to optimize the conditions of synthesis for this particular Al concentration but the transition remained broad for several samples synthesized. Although the x-ray spectra for all compositions show the $AlB_2$ structure as the main phase and the deduced lattice constants change continuously between x=0.6 and x=1.2 (Fig. 1) the broad transition for x=1.1 could not be narrowed within repeated attempts of synthesis. A possible explanation might be a chemical miscibility region leading to a non-uniform distribution of Al and Si in the honeycomb planes.

In order to investigate the effect of vacancies in the sublattices of Ca and Al/Si we have also prepared samples with a slightly higher or lower Ca content. This also has a detrimental effect on the superconducting state. As shown in Fig. 6 the $T_c$ decreases for both, $Ca_{1.1}AlSi$ and $CaAl_{1.1}Si_{1.1}$, as compared to CaAlSi.

According to the phase diagram, the maximum $T_c$ of the ternary compound Ca-Al-Si is obtained for the ideal (1:1:1) composition of Ca, Al, and Si. Any deviation from this 1:1:1 ratio results in a decrease of the superconducting transition temperature. It is, therefore, of interest to compare the effect of vacancies or doping in other ternary systems (e.g. Sr-Ga-Si) on their $T_c$. For $Sr(Ga_{0.37}Si_{0.63})_2$ the recently reported $T_c$ of 3.4 K [2] should increase if the composition is closer to the 1:1:1 ionic ratio. In fact, we have recently observed a maximum of $T_c$ at 5.1 K in SrGaSi [10].

## 4. Conclusions

We have synthesized a series of the pseudo-ternary intermetallic compounds $AAl_{2-x}Si_x$ for A = Ca, Sr and Ba with the C32 structure. The lattice parameters and the c/a ratios of AAlSi increase continuously as the atomic size of A increases from Ca through Sr to Ba. At the same time, $T_c$ decreases from 7.8 K for CaAlSi through 5.1 K for SrAlSi to non-superconducting above 2 K for BaAlSi. The x-ray spectra of $CaAl_{2-x}Si_x$ also reveal a gradual and smooth decrease of the lattice constants with increasing x but an increase of the c/a ratio. The superconducting transition temperature reaches a maximum to 7.8 K for

x=1 and decreases for smaller or larger x values. A possible miscibility region for x slightly larger than 1 is proposed to account for the very broad superconducting transition observed, which shows two steps in the resistivity drop. Changing the ratio of Ca:(Al/Si) also results in a decrease of $T_c$. Based on Seebeck coefficient measurements of the "maximum-$T_c$" samples CaAlSi and SrAlSi we conclude that the electronic structure of this ternary compound is very complex and different from that of $MgB_2$. In particular, there is no simple dependence of $T_c$ on the lattice parameters or the c/a ratio. The maximum $T_c$ for these pseudo-ternary compounds appears at the 1:1:1 ionic composition. A possible doping-induced electronic transition may have occurred in the $CaAl_{2-x}Si_x$ compound system at x ~ 0.75, similar to that in $SrGa_{2-x}Si_x$.

**Acknowledgements**

This work is supported in part by NSF Grant No. DMR-9804325, the T.L.L. Temple Foundation, the John J. and Rebecca Moores Endowment, and the State of Texas through the Texas Center for Superconductivity at the University of Houston and at Lawrence Berkeley Laboratory by the Director, Office of Energy Research, Office of Basic Energy Sciences, Division of Materials Sciences of the U.S. Department of Energy under Contract No. DE-AC03-76SF00098.

Figure Captions

Fig. 1: Lattice parameters of $CaAl_{2-x}Si_x$ as function of x. The lattice constants of SrAlSi are indicated by the asterisks at x=1.

Fig. 2: Seebeck coefficient, S(T), of (a) CaAlSi and (b) SrAlSi. The upper right insets show S(T) near the superconducting transition. Lower left insets display the resistivities near $T_c$.

Fig. 3: Resistance data near the superconducting transition of $CaAl_{2-x}Si_x$ for $0.6 \leq x \leq 1$. The resistance was normalized to its value right above $T_c$.

Fig. 4: Superconducting phase diagram of $CaAl_{2-x}Si_x$. The error bars attached to the data points measure the transition width. At x=1.1 the very broad transition possibly indicates a chemical miscibility gap for this Al:Si ratio.

Fig. 5: The resistive superconducting transition for $CaAl_{0.9}Si_{1.1}$. The broad transition clearly proceeds in two steps due to non-uniform distribution of the Al/Si ions.

Fig. 6: Resistance at the superconducting transitions in $CaAl_{1.1}Si_{1.1}$ and $Ca_{1.1}AlSi$ as compared to CaAlSi.

Fig. 1.

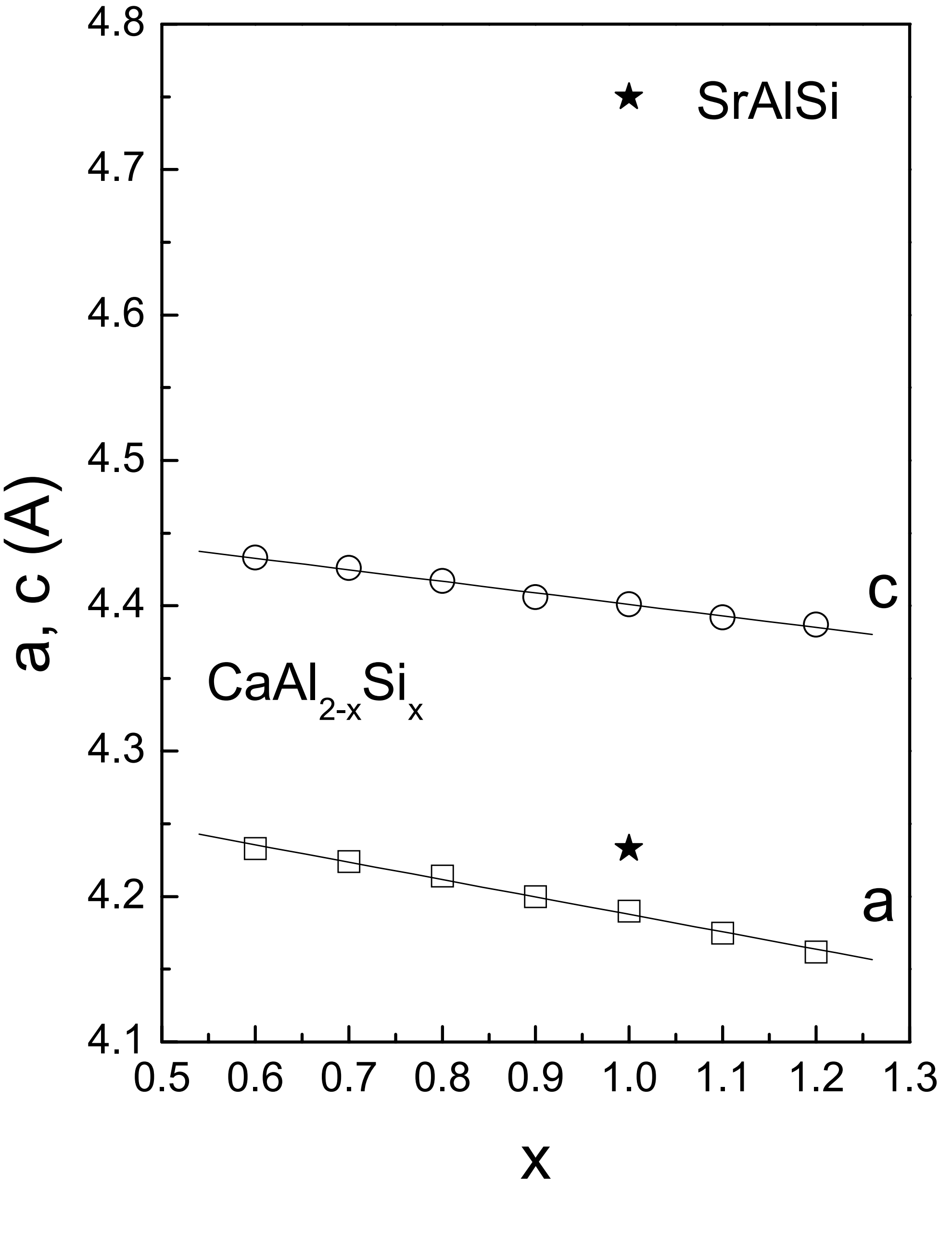

Fig 2a.

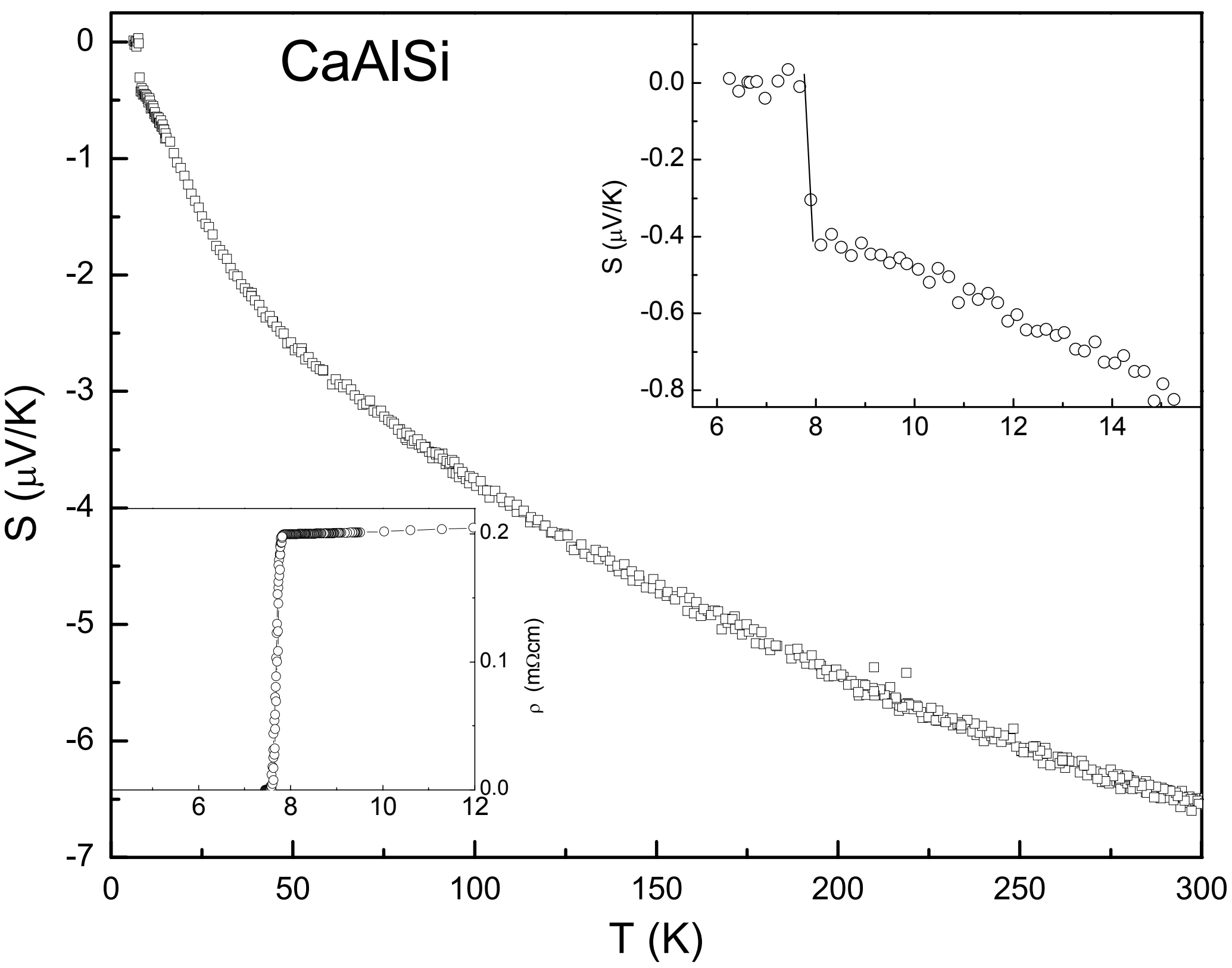

Fig 2b.

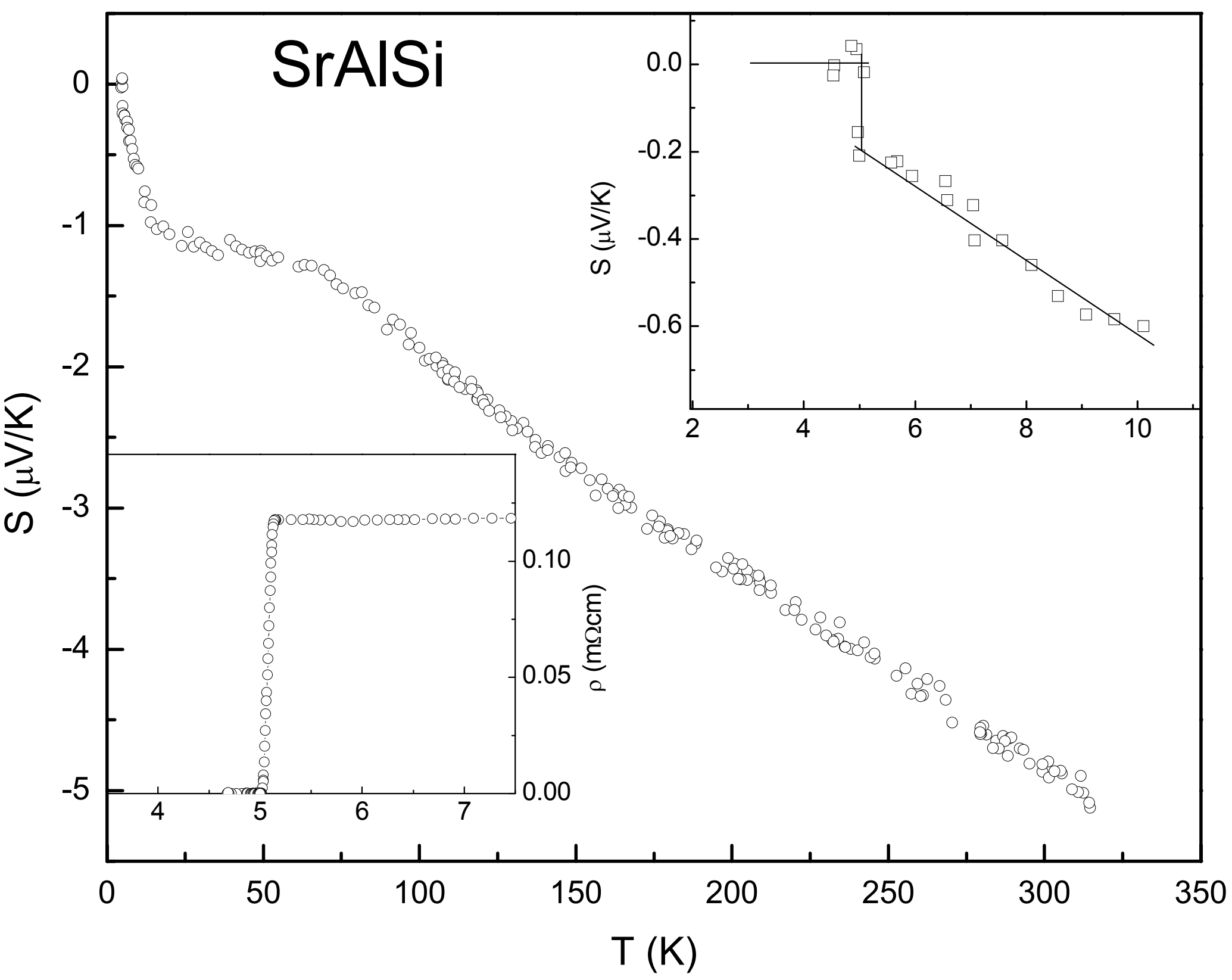

Fig 3.

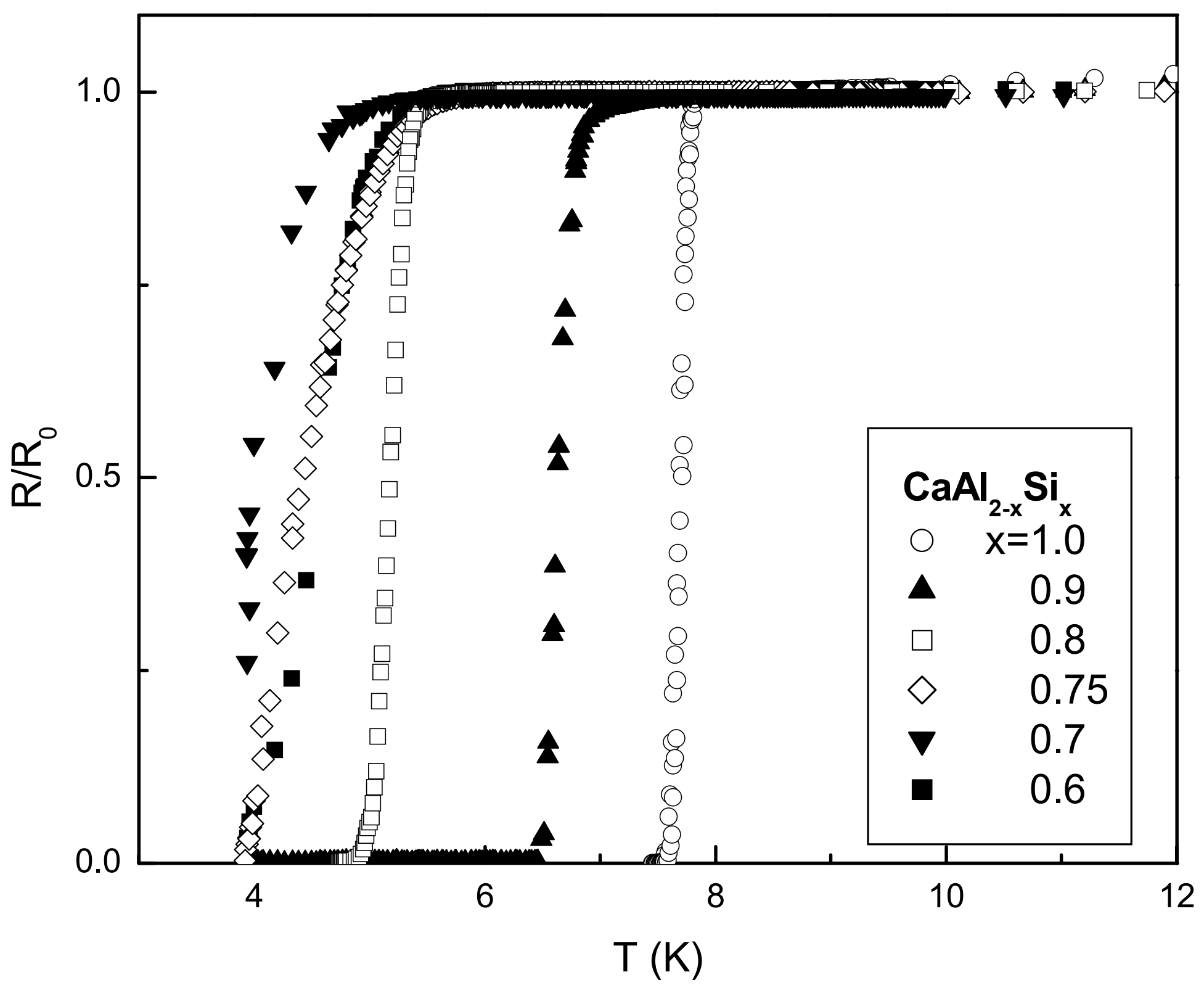

Fig. 4.

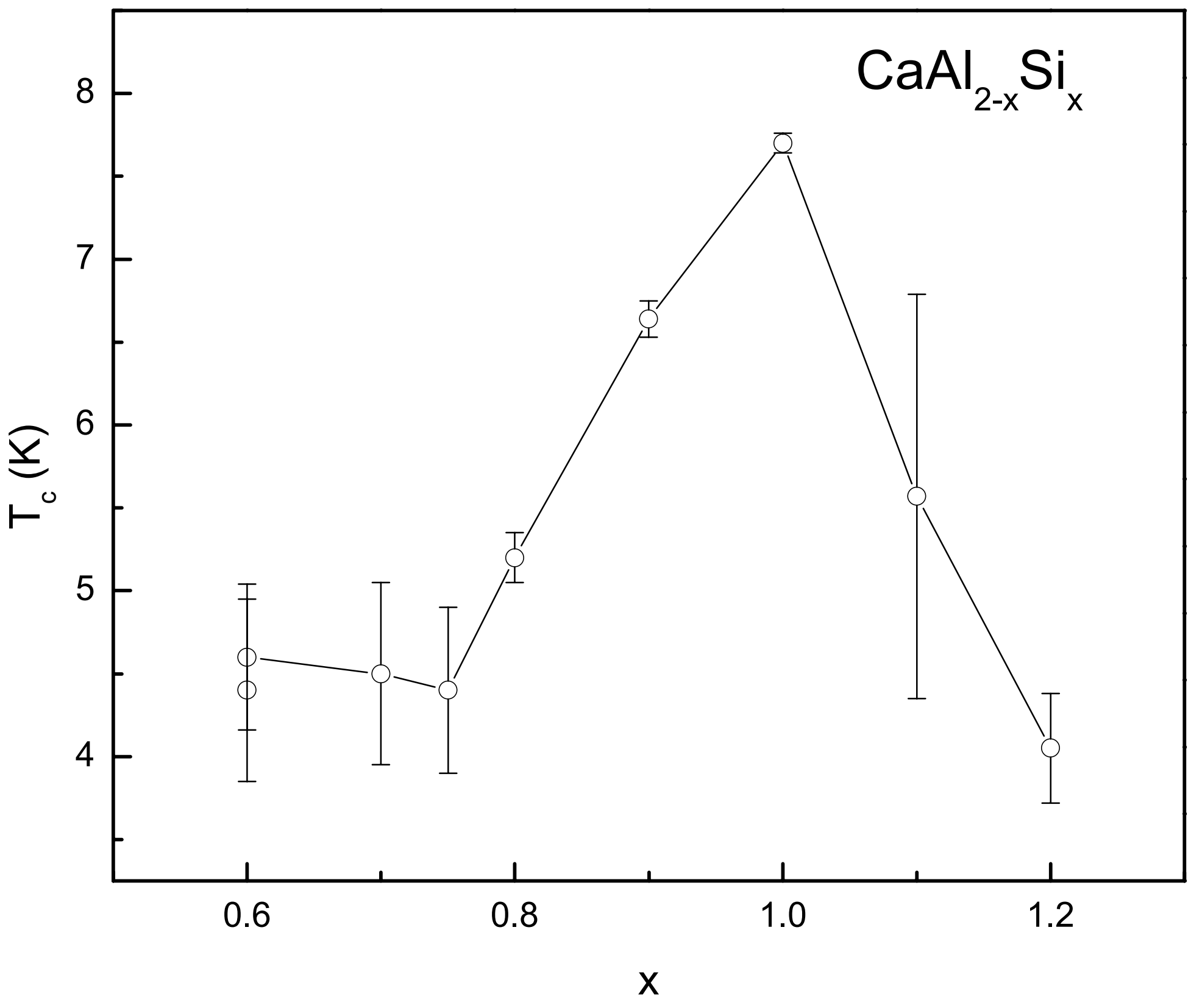

Fig. 5.

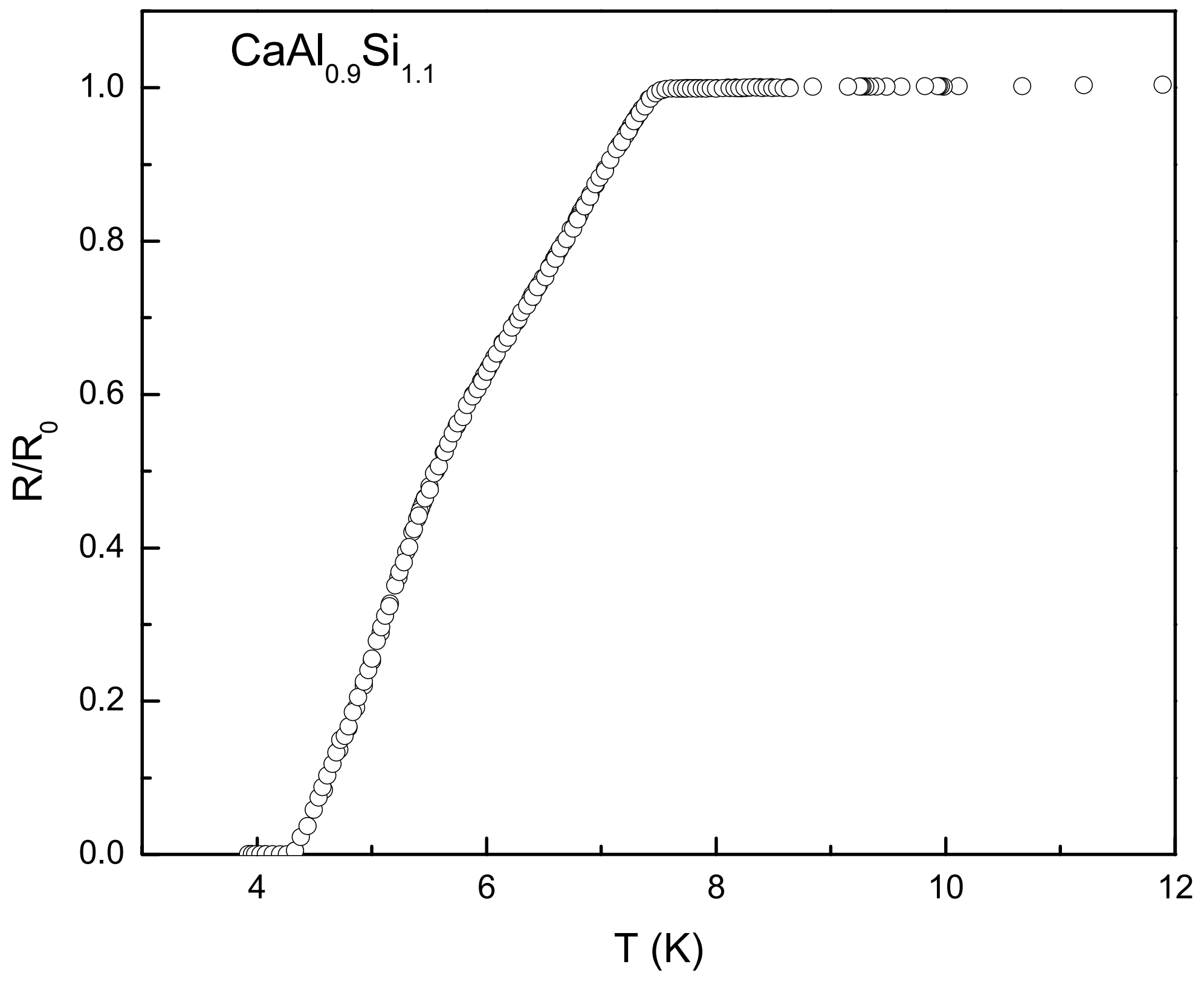

Fig. 6.

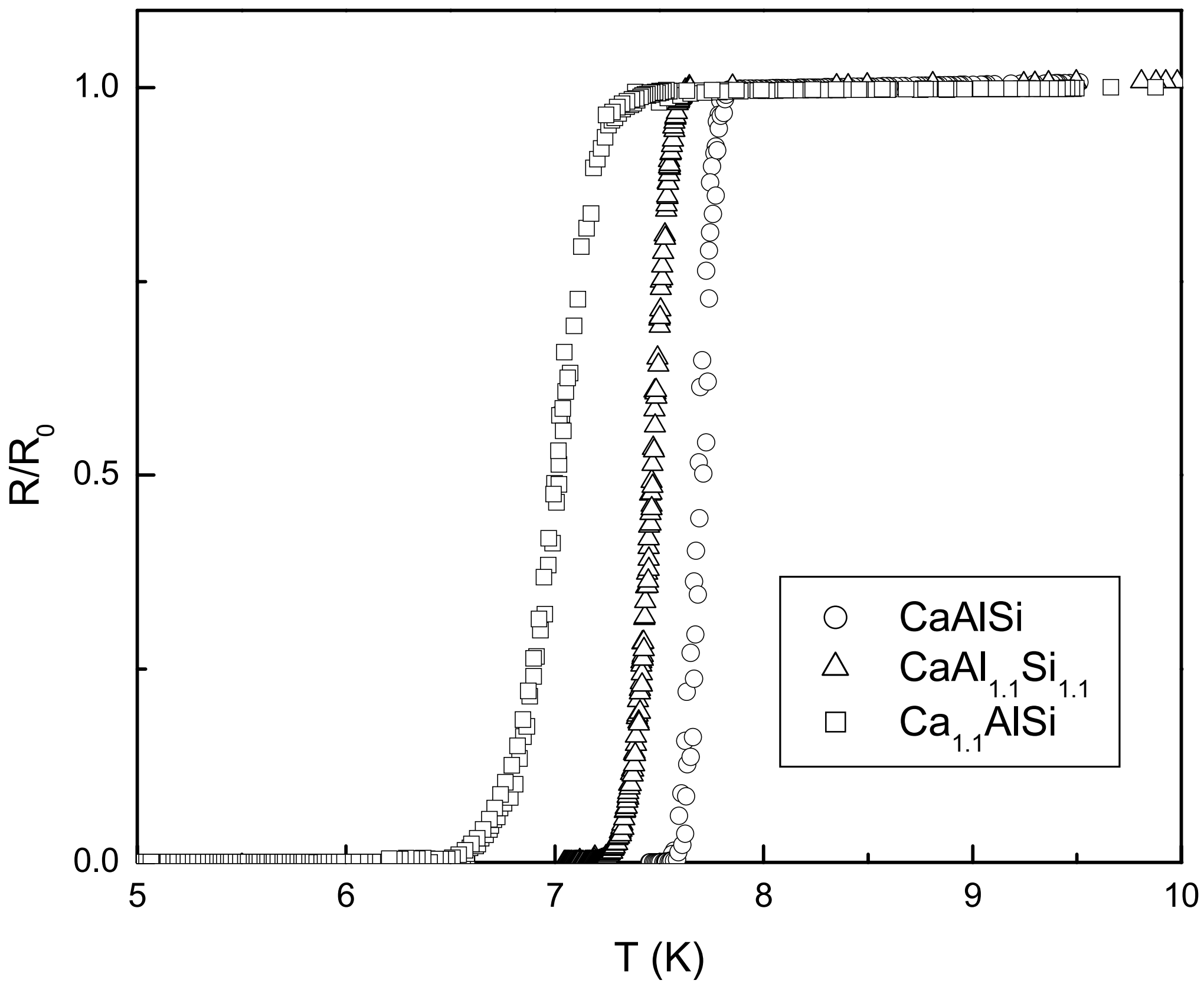